\documentclass[12pt]{article}
\usepackage{epsfig}
\usepackage{graphicx,float}
\usepackage{amsfonts}
\usepackage{amssymb}
\usepackage{arydshln}
\usepackage{latexsym}
\def\ee{\end{equation}}
\def\ba{\begin{eqnarray}}
\def\ea{\end{eqnarray}}

\def\bq{\begin{quote}}
\def\eq{\end{quote}}

 at 10truept

\newcommand{\beq}{\begin{equation}}
\newcommand{\eeq}{\end{equation}}
\newcommand{\beqa}{\begin{eqnarray}}
\newcommand{\eeqa}{\end{eqnarray}}
\newcommand{\bea}{\begin{eqnarray}}
\newcommand{\eea}{\end{eqnarray}}
\newcommand{\p}{\partial}
\newcommand{\al}{\alpha}
 
 \newcommand{\ep}{\epsilon}

\newcommand{\overskrift}[1]{\vspace{6.0mm}\noindent\textbf{#1}\vspace{1.5mm}}

\def\ltap{\ \raise.3ex\hbox{$<$\kern-.75em\lower1ex\hbox{$\sim$}}\ }
\def\gtap{\ \raise.3ex\hbox{$>$\kern-.75em\lower1ex\hbox{$\sim$}}\ }
\def\gl{\ \raise.5ex\hbox{$>$}\kern-.8em\lower.5ex\hbox{$<$}\ }
\def\roughly#1{\raise.3ex\hbox{$#1$\kern-.75em\lower1ex\hbox{$\sim$}}}

\usepackage{amsmath,hyperref,bm}
\newcommand{\od}[2]{ { { #1}^{\scriptscriptstyle(#2)}}}

\newcommand{\ods}[1]{\od{ {#1}}{1}+\od{#1}{2}+\dots}
\newcommand{\RR}{\mathcal{R}_3}

\newcommand{\vev}[1]{\left<0\left|#1\right|0\right>}

\newcommand{\sloep}{ {\epsilon_H}}
\newcommand{\sloet}{ {\eta_H}}
\newcommand{\sloka}{ {\kappa_H}}
\newcommand{\mpla}{ {M_\text{pl}}^2}

\begin{document}
\bibliographystyle{unsrt}
\thispagestyle{empty}
\begin{flushright}
September 17, 2007
\end{flushright}

\vskip2cm
\begin{center}

{\Large{\bf de Sitter limit of inflation}} \\ 
\vskip0.2cm
{\Large{\bf and nonlinear perturbation theory}}

\vskip2cm {\large  Philip R. Jarnhus\footnote{\tt pjarn@phys.au.dk} and Martin S. Sloth\footnote{\tt sloth@phys.au.dk}}\\

\vspace{.5cm}

\vskip 0.1in

{\em Department of Physics and Astronomy, University of Aarhus}\\
{\em DK-8000 Aarhus C, Denmark}\\

\vskip 0.1in
\vskip 0.1in
\vskip .25in
{\bf Abstract}
\end{center}

We study the fourth order action of the comoving curvature perturbation in an inflationary universe in order to understand more systematically the de Sitter limit in nonlinear cosmological perturbation theory. We derive the action of the curvature perturbation to fourth order in the comoving gauge, and show that it vanishes sufficiently fast in the de Sitter limit. By studying the de Sitter limit, we then extrapolate to the $n$'th order action of the comoving curvature perturbation and discuss the slow-roll order of the $n$-point correlation function. 

\vfill \setcounter{page}{0} \setcounter{footnote}{0}
\newpage

\setcounter{equation}{0} \setcounter{footnote}{0}
-

\section{Introduction}

Even though inflation has proven to be a very successful paradigm for the evolution of the early universe, we have very limited knowledge of the theory of inflation itself. At present, even the energy scale at which inflation takes place is unknown within almost 10 orders of magnitude. The cosmological data available has reached a precision, which allow us to test the generic predictions from inflation of a flat, adiabatic and gaussian scalar spectrum of cosmic microwave background (CMB) anisotropies very well \cite{Spergel:2006hy}. On the other hand, the data is not yet precise enough to enable us to convincingly find sub-leading deviations from the generic predictions, which would enable us to discriminate between different classes of inflationary models \cite{Hamann:2006pf}. 

In order to make a transition to the next level in our understanding, it is important to understand more systematically what our theory entails to higher order in perturbation theory. A lot of effort has already been put into understanding second order perturbation theory \cite{Acquaviva:2002ud}, non-gaussianities (for a review see e.g.\cite{Bartolo:2004if}), loop corrections \cite{Weinberg:2005vy,Sloth:2006az,Weinberg:2006ac,Sloth:2006nu,Seery:2007wf,Seery:2007we}, and even general nonlinear perturbation theory \cite{Langlois:2006vv,Enqvist:2006fs}. However, there is one question which is of particular interest to us. We are interested in what we can learn about higher order perturbation theory by studying the de Sitter limit. Therefore, in order to make a step in this direction, in this paper we will study the de Sitter limit of cosmological perturbations up to fourth order in two different gauges; the comoving curvature gauge and the uniform curvature gauge. 

The action of cosmological perturbations to third order has previously been calculated in both gauges in ref.~\cite{Maldacena:2002vr}, and later generalized to include higher dimensional operators \cite{Creminelli:2003iq,Seery:2005wm}, and multi field scenarios \cite{Seery:2005gb}. It was found that the third order terms in the action are slow-roll suppressed in both gauges and thus vanishes in the de Sitter limit. This is expected, since the curvature perturbations should be a pure gauge in the de Sitter limit, and no longer be dynamical. One may expect that in the uniform curvature gauge the action to all orders higher than two should be slow-roll suppressed in the pure de Sitter limit \cite{Seery:2005gb}, but an explicit calculation of the fourth order action in the uniform curvature gauge shows that the action of inflaton fluctuations is {\it not} slow-roll suppressed at fourth order \cite{Sloth:2006az,Seery:2006vu}.  As discussed in \cite{Seery:2006vu}, this happens because the gravity is highly nonlinear and the nonlinearity in the inflaton fluctuations are controlled by the strength of the gravitational interaction rather than the slow-roll parameter.  However, when the action is reformulated in terms of the comoving curvature perturbation, it becomes slow-roll suppressed. This confirms that the curvature perturbation in pure de Sitter space is a pure gauge, which can also be understood from the fact that the transformation from inflaton perturbations into comoving curvature perturbations becomes singular in the pure de Sitter limit. These are the issues we will explore in more explicit details in the present paper.

In the next section, we calculate explicitly the fourth order action of the comoving curvature perturbation. This is the main result of the paper. We show that by a direct application of the ADM formalism, one obtains a result which is zeroth order in the slow-roll expansion. Just like in the third order calculation \cite{Maldacena:2002vr}, the zeroth order terms can be eliminated by partial integrations to give an action, which is suppressed by one power of the slow-roll parameter. However, the leading contribution is proportional to the linear perturbation equation, and can be eliminated by a change of variables, leading finally to an action, which is suppressed by two powers of the slow-roll parameters.  

By studying the de Sitter limit we extrapolate our findings from fourth order to $n$'th order. We argue that the action of inflaton perturbations in the uniform curvature gauge has to be slow-roll suppressed to any odd power, while it will remain unsuppressed to any even power in the perturbation expansion.  We then  review systematically the slow-roll order of $n$-point correlation functions of the comoving curvature perturbations up to $n=4$, and discuss the extrapolation to any $n>4$.  As an example, we conjecture that the nonlinearity parameters of the $5$- and $6$-point functions are $\mathcal{O}(\ep^2)$, where $\ep$ is the first slow-roll parameter. The nonlinearity parameters of the $3$- and $4$-point functions, $f_{NL}$, $\tau_{NL}$, have previously been shown to be of order $\mathcal{O}(\ep)$ \cite{Maldacena:2002vr,Seery:2006vu}. The results for the nonlinearity parameter to $n$'th order is summarized in table 2 at the end of section 3.

Although our motivation for deriving the action to fourth order in the comoving gauge is more conceptual, the result may also have more phenomenological applications. It can be used for calculating loop effects\footnote{In ref.~\cite{Sloth:2006az,Sloth:2006nu} the fourth order action in the uniform curvature gauge was used to demonstrate that the loop corrections to inflation can be large, and may have observationally important consequences for the predicted tensor-to-scalar relation in models of chaotic inflation. This is consistent with ref.~\cite{Weinberg:2005vy,Weinberg:2006ac}, and has later been verified using a combination of the fourth order action in the uniform curvature gauge and the $\delta N$-approach \cite{Seery:2007wf,Seery:2007we}. It is possible to make use of the $\delta N$-approach, because the IR contributions to the loops also can be described by a particular classical approximation \cite{van der Meulen:2007ah} (see also \cite{Musso:2006pt}).}, and non-adiabatic enhancements of the tri-spectrum beyond the slow-roll approximation \cite{hhjs}.

The outline of the paper is the following. In section 2 we derive the fourth order action of curvature perturbations in the comoving gauge. In section 3 we discuss the de Sitter limit and the extrapolation of the results to arbitrary order. In section 4 we summarize and conclude on our findings.


\section{The action to fourth order in comoving gauge}

The action of cosmological perturbations during inflation has previously been calculated both in the comoving and in the uniform curvature gauge to third order, but only in the uniform curvature gauge to fourth order. The two gauges have different benefits, and it is useful to complete the picture to fourth order. Thus, we proceed by computing the action of curvature perturbations to fourth order in the comoving gauge.

In order to calculate the action, it is convenient to use the ADM formalism \cite{Arnowitt:1962hi}, and slice the 4-geometry into a sandwich structure along the time-coordinate. One can produce a series of spatial 3-geometries of uniform time, separated by time steps $\mathrm{d}t$. To connect two spatial slices, one introduces the lapse function, $N$. Similarly one can write the difference between the position in the two 3-geometries in terms of the shift-vector $N^{i}$ \cite{Arnowitt:1962hi}
\begin{equation}
    x^i_{\text{new}}=x^i_{\text{initial}}-N^i\mathrm{d}t~.
    \label{shift}
\end{equation}
The quantities $N$ and $N^i$ will later be expanded in series of the perturbation parameters. With these definitions, one can write the line element on the ADM form \cite{Arnowitt:1962hi}, in terms of $N$, $N^i$, 
\begin{equation}
            ds^2=-N^2dt^2+h_{ij}(dx^i+N^idt)(dx^j+N^jdt),
        \label{ADM}
\end{equation}
with $h_{ij}$ being a purely spatial metric. Using this, the Einstein-Hilbert action for a single scalar field
\begin{equation}
    S=\int \textrm{d}^4x\sqrt{-g}\left[ R+\frac{1}{2}\partial_\mu\phi\partial^\mu\phi-V(\phi) \right]~,
    \label{genaction}
\end{equation}
can be written as
\begin{equation}
        \begin{split}
                S=\frac{1}{2}\int\textrm{d}t\textrm{d}^3x\sqrt{h}\Big[&N\RR-2NV(\phi)+N^{-1}(E_{ij}E^{ij}-E^2)+N^{-1}(\dot\phi-N^i\partial_i\phi)^2\\ 
                &-Nh^{ij}\partial_i\phi\partial_j\phi\Big]
        \end{split}
        \label{ADMaction}
\end{equation}
with $\RR$ being the curvature scalar associated with $h_{ij}$. The tensor $E_{ij}$, which is closely related to the extrinsic curvature, is defined as a linear combination of the time derivative of $h_{ij}$ and the covariant derivative of the shift vector
\begin{equation}
        E_{ij}=\frac{1}{2}(\dot h_{ij}-\nabla_iN_j-\nabla_jN_i)~,
        \label{Eij}
\end{equation}
and $E\equiv E^i_i$ is the trace of the tensor $E_{ij}$. 


\subsection{Gauge choice, constraints, and solutions}

In this and the following subsection, we will work in the comoving curvature gauge, defined as
\begin{equation} 
    \phi=\phi_c(t)\quad h_{ij}=a(t)^2(e^{2\zeta(\bm{x},t)}\delta_{ij}+\gamma_{ij})~,~\gamma_{ii}=0\quad\partial_i\gamma_{ij}=0. 
         \label{gauge}
\end{equation}
For the purpose of this paper, we can focus on scalar perturbations of the metric, thus letting $\gamma_{ij}$ being identical to 0.

In the present gauge the curvature scalar becomes $\RR=-4\partial^2\zeta-2\left( \partial\zeta \right)^2$\footnote{We use the notation $\partial^2\zeta=h^{ij}\partial_i\partial_j\zeta$ and $\left( \partial\zeta \right)^2=h^{ij}\partial_i\zeta\partial_j\zeta$.}, and from the action in eq.~(\ref{ADMaction}), one can derive the equations of motion $N^i$, 
\begin{equation}
         \nabla_i\left[ N^{-1}(E^i_j-\delta^i_jE) \right]=0~,
     \label{Const.1}
\end{equation}
and for $N$
\begin{equation}
        \RR-2V-N^{-2}(E_{ij}E^{ij}-E^2)-N^{-2}{\dot\phi}^2=0~.
        \label{Const.2}
\end{equation}
If one furthermore expands the scalar and shift parameter in terms of three perturbation parameters to an arbitrary order,  one obtains
\begin{equation}
        N=1+(\ods{\alpha})\quad N_i=\partial_i(\ods\chi)+(\ods{\beta_i})
        \label{perturb}
\end{equation}
with the constraint (cf. Helmholtz theorem)
\begin{equation}
        \partial^i\beta_i=0~.
        \label{betadiv}
\end{equation}
It is now possible to solve equations (\ref{Const.1}) and (\ref{Const.2}) order by order for all three perturbation parameters. As it will turn out, we only need to calculate the first and second order perturbation parameters. The fourth order terms only appears in the action multiplied by the equation of motion for the background field, $\phi_c$, while the third order terms cancel one another. From eq.~(\ref{Const.1}), by taking the divergence and employing eq.~(\ref{betadiv}), one finds to first order in $\alpha$  \cite{Maldacena:2002vr},
\begin{equation}
        \od{\alpha}{1}=H^{-1}\dot\zeta~.
        \label{alpha1}
\end{equation}
Inserting this back into the first order contribution to eq.~(\ref{Const.1}), and solving for the first order vector perturbation, one also obtains  \cite{Maldacena:2002vr}
\begin{equation}
        \od{\beta_i}{1}=0~.
        \label{beta1}
\end{equation}
In eq.~(\ref{alpha1}) we have introduced the parameter $H\equiv\dot a /a$. 

Utilizing the same techniques for the second order contributions, we get
\begin{equation}
    \begin{split}
        \od{\alpha}{2}=\frac{1}{2H}\partial^{-2}\Big\{ &\partial^j\left( \partial_j\od{\alpha}{1}\partial^2\od{\chi}{1}-\partial_i\od{\alpha}{1}\partial^i\partial_j\od{\chi}{1} \right)-2\partial_i\partial_j\left( \partial^j\od{\chi}{1}\partial^i\zeta \right)\\
                &+\partial^j\left( \partial_l\zeta\partial^l\partial_j\od{\chi}{1} \right)+\partial^j\left( \partial_j\zeta\partial^2\od{\chi}{1} \right) \Big\}
        \label{alpha2}
\end{split}
\end{equation}
for the scalar part, and
\begin{equation}
    \begin{split}
        \od{\beta_j}{2}=2\partial^{-2}\Big\{& \partial_j\od{\alpha}{1}\partial^2\od{\chi}{1}-\partial_i\od{\alpha}{1}\partial^i\partial_j\od{\chi}{1}+\partial_j\zeta\partial^2\od{\chi}{1}-2H\partial_j\od{\alpha}{2}\\ 
        &-\partial_i\left( \partial_j\od{\chi}{1}\partial^i\zeta+\partial^i\od{\chi}{1}\partial_j\zeta \right)+\partial_l\zeta\partial^l\partial_j\od{\chi}{1}\Big\}
    \end{split}
        \label{beta2}
\end{equation}
for the vector perturbation. In the equations above $\od{\chi}{1}$ is the first order part of the third perturbation parameter, which we will solve for in a moment. From this we see, that $\alpha$ to the lowest order only describe how the metric changes in time, as could be expected from its role in the metric.

Similarly we can solve eq.~(\ref{Const.2}) for the $\chi$ parameters, whereby one gets 
\begin{equation}
        \od{\chi}{1}=-\frac{\zeta}{H}+\xi\quad,\quad \partial^2\xi=\frac{H^{-2}}{2}\dot\phi^2\dot\zeta
        \label{chi1}
\end{equation}
to the first order \cite{Maldacena:2002vr}. Likewise to the second order in $\chi$, we obtain (for readability we refrain from inserting the values of the $\alpha$ and $\beta$)
\begin{equation}
    \begin{split}
        4H\partial^2\od{\chi}{2}=&-2\left( \partial\zeta \right)^2+2(\dot\phi^2-6H^2)\od{\alpha}{2}+4\dot\zeta\partial^2\od{\chi}{1}+4\dot\phi^2\al^{(1)}\zeta\\ 
                &-4H\partial_l\od{\chi}{1}\partial^l\zeta-3\od{\alpha}{1}^2\dot\phi^2+\partial^2\od{\chi}{1}\partial^2\od{\chi}{1}-\partial_i\partial_j\od{\chi}{1}\partial^i\partial^j\od{\chi}{1}
        \label{chi2}
\end{split}
\end{equation}

It is seen, that the $\beta$ and $\chi$ parameters of a given order depends on the $\alpha$ parameter of the same order, as opposed to the $\alpha$ that only depends on parameters of lower orders. This translates into a dependence of the lapse function in the shift parameter, which is reasonable as the lapse function is linked to the thickness of the time-slices, and in that way affect the size of the shift.


\subsection{Fourth order action}
Truncating the action at the fourth order and exploiting the equation of motion for the background field
\begin{equation}
    \dot\phi^2-6H^2+2V=0~,
    \label{0ordcon}
\end{equation}
we write the action in terms of the perturbation parameters, and simplify it by using partial spatial integrations. This gives
\begin{equation}
	\begin{split}
		\od{S}{4}=\frac{1}{2}\int \textrm{d}t\textrm{d}^3x a^3\Bigg\{&-\frac{1}{3}\zeta^3\p^2\zeta-2\od{\alpha}{1}(\zeta\p_i\zeta\p^i\zeta+\zeta^2\p^2\zeta)+\dot\phi_c^2\od{\alpha}{1}^2\left[ \frac{9}{2}\zeta^2-3\zeta\od{\alpha}{1}+\od{\alpha}{1}^2 \right]\\
				&\left[\frac{1}{2}\zeta^2+\zeta\od{\alpha}{1}+\od{\alpha}{1}^2\right]\left[ \partial_i\partial_j\od{\chi}{1}\partial^i\partial^j\od{\chi}{1}-\partial^2\od{\chi}{1}\partial^2\od{\chi}{1} \right]+(6H^2-\dot\phi^2)\od{\alpha}{2}^2\\
		&-2[\zeta+\od{\alpha}{1}]\left[ \partial_i\partial_j\od{\chi}{1}\partial^i\partial^j\od{\chi}{2}-\partial^2\od{\chi}{1}\partial^2\od{\chi}{2}-2\p_i\p_j\od{\chi}{1}\p^i\od{\chi}{1}\p^j\zeta\right]\\
		&-2\left[ 2\partial_i\p_j\od{\chi}{2}\partial^i\od{\chi}{1}\partial^j\zeta+2\partial_i\p_j\od{\chi}{1}\partial^i\od{\chi}{2}\partial^j\zeta- \p_j\od{\chi}{1}\p_i\zeta\partial^i\od{\chi}{1}\partial^j\zeta \right]\\
		&+\frac{1}{2}\partial_i\od{\beta_j}{2}\partial^i\od{\beta^j}{2}-2\od{\alpha}{1}\partial_i\partial_j\od{\chi}{1}\partial^i\od{\beta^j}{2}\Bigg\}
	\end{split}
	\label{4ordaction}
\end{equation}
Though fairly compact in its presentation, it is not clear from this form that the action in eq.~(\ref{4ordaction}) does indeed vanish in the slow roll limit. The zeroth order slow-roll terms can however be removed by partial integrations. To make this more evident, it is simpler to proceed with finding a gauge transformation between the uniform curvature gauge and the present gauge, in order to calculate the action by a gauge transformation of the equivalent action in the uniform curvature gauge.


\subsection{Gauge transformation}

One can find the gauge transformation by doing a translation in the time coordinate in order to transform between the uniform curvature gauge and the comoving gauge. Below we follow a procedure, which is similar to the one in ref.~\cite{Maldacena:2002vr}. The uniform curvature gauge is given by
\begin{equation}   \label{isocurvgauge}
    \phi(t,\bm{x})=\phi_c(t)+\delta\phi(t,\bm{x})~,~h_{ij}=a(t)^2\delta_{ij}~,
  \end{equation}
where we have defined the background field $\phi_c\equiv\vev{\phi}$, such that the tadpole condition, $\vev{\delta\phi}=0$, is satisfied.

Choosing the time translation given by the vector $\xi_\mu=(T,0,0,0)$, we can write the transformed field fluctuation \cite{Bruni:1996im}
\begin{equation}
	\begin{split}
		\delta\phi(x_\mu+\xi_\mu)=\delta\phi(x_\mu)+\sum_{n=1}^\infty\frac{(\xi_\mu\partial^\mu)^n}{n!}\phi(x_\mu)~.
	\end{split}
	\label{Lie}
\end{equation}
The sum is nothing more than the Taylor expansion along a vector. Requirering the time translation to bring the coordinates to the comoving gauge, we fix $\xi_\mu$ by $\delta\phi(x_\mu+\xi_\mu)=0$. Evolving order by order we then get for $\xi_0=\ods{T}$.
\begin{align}
	\od{T}{1}&=-\frac{\delta\phi}{\dot\phi_c}\\
	\od{T}{2}&=\frac{\delta\phi\dot{\delta\phi}}{ {\dot\phi_c}^2}-\frac{1}{2}\frac{\ddot\phi_c}{ {\dot\phi_c}^3}\delta\phi^2\\
	\od{T}{3}&=\frac{3}{2}\frac{\ddot\phi_c}{ {\dot\phi_c}^4}\delta\phi^2\dot{\delta\phi}-\frac{1}{2}\frac{\delta\phi^2}{ {\dot\phi_c}^3}\ddot{\delta\phi}-\frac{\delta\phi\dot{\delta\phi}^2}{ {\dot\phi_c}^3}-\left( \frac{1}{2}\frac{\ddot{\phi_c}^2}{ {\dot\phi_c}^5}-\frac{1}{6}\frac{\dddot\phi_c}{ {\dot\phi_c}^4} \right)\delta\phi^3
	\label{Torders}
\end{align}
As a method for finding the relation between $\zeta$ in the comoving gauge and $\delta\phi$ in the uniform curvature gauge, we perform coordinate transformation above in the metric associated with the uniform curvature gauge ($g_{\mu\nu}^{(\textrm{uc})}$) and equate it with metric in the comoving gauge ($g_{\mu\nu}^{(\textrm{cm})}$). 

The spatial part of the metric gives after the time translation (again ignoring tensor contributions)
\begin{equation}
	h_{ij}^{(\textrm{cm})}=a^2(t+T)\delta_{ij}-N^2\partial_iT\partial_jT+\partial_iTN_j+\partial_jTN_i~,
	\label{metricredef}
\end{equation}
with $N$ and $N_i$ defined as in eq.~(\ref{perturb}). Solving the first order contribution in the above equation gives $\zeta=HT$ or \cite{Maldacena:2002vr}
\begin{equation}
	\zeta=-H\frac{\delta\phi}{\dot\phi_c}\equiv\zeta_n~.
	\label{zetan}
\end{equation}
No spatial reparametrization is needed for the first order. To higher orders one must do a transformation $x_i\to x_i+\upsilon_i$, such that
\begin{equation}
-N^2\partial_iT\partial_jT+(\partial_iTN_j+\partial_jTN_i)+\partial_i\upsilon_j+\partial_j\upsilon_i=\exp\left(2\sum_n\iota_n\right)a^2(t)\eta_{ij}
	\label{reparam}
\end{equation}
with $\iota_n$ being a parameter of the $n$'th order. Recalling that $\iota_1=0$ and writing $a(t)=e^{\rho(t)}$ like in ref.~\cite{Maldacena:2002vr}, we can write the equation for $\zeta$ up to third order
\begin{equation}
	\zeta=\rho(t+T)-\rho(t)+(\iota_2+\iota_3)
	\label{zetaneq}
\end{equation}
By taking the trace and $\partial^i\partial^j$ of eq.~(\ref{reparam}), one can solve to second order
\begin{equation}
	4\iota_2=2\partial^i\od{T}{1}\partial_i\od{\chi_\phi}{1}-2\partial^{-2}\partial^i\partial^j(\partial_i\od{T}{1}\partial_j\od{\chi_\phi}{1})-(\partial_i\od{T}{1}\partial^i\od{T}{1}-\partial^{-2}\partial^i\partial^j(\partial_i\od{T}{1}\partial_j\od{T}{1}))
	\label{iota2}
\end{equation}
as $\od{\beta_i}{1}=0$. Thereby giving \cite{Maldacena:2002vr}
\begin{equation}
	\begin{split}
		\zeta=\zeta_n-f_2(\zeta_n)=&\zeta_n+\frac{1}{2}\frac{\ddot\phi_c}{\dot\phi_cH}\zeta_n^2+\frac{1}{4}\frac{\dot\phi_c^2}{H^2}\zeta_n^2+\frac{\zeta_n\dot\zeta_n}{H}+\frac{1}{2}\partial^i\zeta_n\partial_i\od{\chi_\phi}{1}-\frac{1}{2}\partial^{-2}\partial^i\partial^j(\partial_i\zeta_n\partial_j\od{\chi_\phi}{1})\\
		&-\frac{1}{4H^2}\left( \partial_i\zeta_n\partial^i\zeta_n-\partial^{-2}\partial^i\partial^j(\partial_i\zeta_n\partial_j\zeta_n) \right)
	\end{split}
	\label{zeta2ndord}
\end{equation}
where we implicitly defined $f_2(\zeta_n)$. We refer to appendix \ref{aux} for the definitions of the functions of the type $\al_\phi^{(i)}$, $\chi_{\phi}^{(i)}$, and $\beta_{\phi}^{(i)}$.
Computing the third order terms in a similar fashion, we get
\begin{equation}
	\begin{split}
		2\iota_3=&\partial^i\od{T}{1}\partial_i\od{\chi_\phi}{2}+\partial^i\od{T}{1}\od{ {\beta_\phi}_i}{2}+\partial^i\od{T}{2}\partial_i\od{\chi_\phi}{1}-\partial^{-2}\partial^i\partial^j(\partial_i\od{T}{1}\partial_j\od{\chi_\phi}{2})\\
		&-\partial^{-2}\partial^j(\partial^i\partial_j\od{T}{1}\od{ {\beta_\phi}_i}{2})-\partial^{-2}\partial^i\partial^j(\partial_i\od{T}{2}\partial_j\od{\chi_\phi}{1})\\
		&-[\partial_i\od{T}{1}\partial^i\od{T}{2}-\partial^{-2}\partial^i\partial^j(\partial_i\od{T}{1}\partial_j\od{T}{2})]\\
		&-\od{\alpha_\phi}{1}[\partial_i\od{T}{1}\partial^i\od{T}{1}-\partial^{-2}\partial^i\partial^j(\partial_i\od{T}{1}\partial_j\od{T}{1})]
	\end{split}
	\label{iota3}
\end{equation}
Repeating the calculation leading to eq.~(\ref{zeta2ndord}), we arrive at
\begin{equation}
	\begin{split}
		\zeta=&\zeta_n-f_2(\zeta_n)-f_3(\zeta_n)\\ =&\zeta_n-f_2(\zeta_n)\\
		&+\frac{5}{6}\frac{\dot\phi_c\ddot\phi_c}{H^3}\zeta_n^3+\frac{1}{3}\frac{\dddot\phi_c}{\dot\phi_cH^2}\zeta_n^3+\frac{1}{4}\frac{\dot\phi_c^4}{H^4}\zeta_n^3+\frac{3}{2}\frac{\ddot\phi_c}{\dot\phi_cH}\zeta_n^2\dot\zeta_n+\frac{\dot\phi_c^2}{H^3}\zeta_n^2\dot\zeta_n+\frac{\zeta_n\dot\zeta_n^2}{H^2}\\
                &+\frac{1}{2}\frac{\zeta_n^2\ddot\zeta_n}{H^2}+\frac{1}{2H}\partial^i\zeta_n\partial_i\od{\chi_\phi}{2}+\frac{1}{2H}\partial^i\zeta_n\od{ {\beta_\phi}_i}{2}\\
                &+\frac{1}{2}\left[ \frac{\ddot\phi_c}{\dot\phi_cH^2}\zeta_n+H^{-1}\dot\zeta_n+\frac{\dot\phi_c^2}{H^2}\zeta_n \right]\left[ \frac{\dot\phi_c^2}{2H^2}\partial^i\zeta_n\partial_i\partial^{-2}\dot\zeta_n-H^{-1}\partial^{i}\zeta_n\partial_i\zeta_n \right]\\
                &+\frac{\zeta_n}{2H^2}\left( \frac{\dot\phi_c^2}{2H^2}\partial^i\dot\zeta_n\partial_i\partial^{-2}\dot\zeta_n-H^{-1}\partial^i\dot\zeta_n\partial_i\zeta_n \right)-\frac{1}{2H}\partial^{-2}\partial^i\partial^j\left( \partial_i\zeta_n\partial_j\od{\chi_\phi}{2} \right)\\                                                                           &-\frac{\partial^{-2}\partial^i\partial^j}{2}\left\{\left[ \frac{\ddot\phi_c}{\dot\phi_cH^2}\zeta_n+H^{-1}\dot\zeta_n+\frac{\dot\phi_c^2}{H^2}\zeta_n \right]\left[ \frac{\dot\phi_c^2}{2H^2}\partial_i\zeta_n\partial_j\partial^{-2}\dot\zeta_n-H^{-1}\partial_{i}\zeta_n\partial_j\zeta_n \right]\right\}\\
                &-\partial^{-2}\partial^i\partial^j\left[\frac{\zeta_n}{2H^2}\left( \frac{\dot\phi_c^2}{2H^2}\partial_i\dot\zeta_n\partial_j\partial^{-2}\dot\zeta_n-H^{-1}\partial_i\dot\zeta_n\partial_j\zeta_n \right)\right]-\frac{1}{2H}\partial^{-2}\partial^j(\partial^i\partial_j\zeta_n\od{ {\beta_{\phi}}_i}{2})\\
                &+\frac{\dot\phi_c^2\zeta_n}{4H^4}\left[ \partial_i\zeta_n\partial^i\zeta_n-\partial^{-2}\partial^i\partial^j(\partial_i\zeta_n\partial_j\zeta_n) \right]~,
	\end{split}
	\label{zeta3rdord}
\end{equation}
where we have implicitly defined the third order contribution in terms of $f_3(\zeta_n)$. Again we refer to appendix \ref{aux} for the definition of functions of the type $\al_\phi^{(i)}$, $\chi_{\phi}^{(i)}$, and $\beta_{\phi}^{(i)}$.

By inverting eq.~(\ref{zeta3rdord}) to third order and recalling the definition of $\zeta_n$ from eq.~(\ref{zetan}), we are able to calculate the fourth order action of the metric perturbations in the comoving gauge from the action of the inflaton field fluctuations in the uniform curvature gauge. Since the calculation is trivial but tedious, we will not repeat it here in all details. However, from the calculation, one would obtain an action for $\zeta$ which is suppressed by only one power of the slow-roll parameter. This can be seen from the gauge transformation above, which contains terms in $f_3$ of order $\zeta_n^3$, but with no slow-roll suppression. To third order in the gauge transformation we can just replace $\zeta_n$ with $\zeta$ in $f_3$, and applied in the second order action for $\delta\phi$ we see immediately that the fourth order action in $\zeta$ will receive contributions, which are only suppressed by one slow-roll order. From this one could easily be led to believe, that the trispectrum of the metric perturbation $\zeta$ is of the wrong order in the slow-roll parameter $\ep$. 

However, since all the leading order slow-roll terms comes from using $f_3$ in the second order action in $\delta\phi$, by partial integrations they can all be rewritten as terms in the action proportional to the linear equation of motion of perturbations. This complication can be avoided by choosing the gauge of $\zeta_n$, then calculate $\left<\zeta_n^4\right>$, and afterwards obtaining the super-horizon value of $\left<\zeta^4\right>$ by virtue of the super-horizon limit of eq.~(\ref{zeta3rdord}), analogous to the calculation in ref.~\cite{Maldacena:2002vr}.  Once the terms proportional to the linear perturbation equation of motion have been eliminated by a change of variables $\zeta\to\zeta_n$, one obtains an action in $\zeta_n$, which is suppressed by two powers of the slow-roll parameter. The full action in the new variable $\zeta_n$ has the following form
\begin{equation}
	\begin{split}
		S_{\zeta_n}=\int\textrm{d}t\textrm{d}^3xa^3\Bigg\{&-\frac{\sloep^2}{6}V''''\zeta_n^4+\frac{1}{2}\partial_i\od{ {\beta_\phi}_j}{2}\partial^i\od{ {\beta_\phi}^j}{2}+\sloep^3\partial_i\partial^{-2}\dot\zeta_n\partial_j\partial^{-2}\dot\zeta_n\partial^i\zeta_n\partial^j\zeta_n\\
		&-2\sloep\left[ -H\sloet\zeta_n+\dot\zeta_n \right]\left[ \partial_i\od{\chi_\phi}{2}+\od{ {\beta_\phi}_i}{2} \right]\partial^i\zeta_n\\
		&+\od{\alpha_\phi}{2}\left[ -6H^2+\dot\phi_c^2 \right]\left[ \sloep^2\zeta_n^2-\frac{\od{\alpha_\phi}{2}}{2} \right]\\
		&-\sloep^2\zeta_n\Big[\big(\frac{\dot\phi_c}{3H}V'''+2H^2\eta_H\ep_H^2-H^2\kappa_H\ep_H+3H^2\ep_H(\eta_H-\ep_H)\big)\zeta_n^3\\
		&+\partial^2\od{\chi_\phi}{2}\dot\zeta_n+\sloep\zeta_n\partial_i\zeta_n\partial^i\zeta_n-\partial_i\partial_j\od{\chi_\phi}{2}\partial^i\partial^j\partial^{-2}\dot\zeta_n\\
		&-\partial_i\od{ {\beta_\phi}_j}{2}\partial^i\partial^j\partial^{-2}\dot\zeta_n+2\sloep(2(\sloep-\sloet)H\zeta_n+\dot\zeta_n)\partial_i\zeta_n\partial^i\partial^{-2}\dot\zeta_n\Big]\Bigg\}~,
	\end{split}
	\label{zetanaction}
\end{equation}
where we have introduced the slow roll parameters associated with the Hamilton-Jacobi formalism:
\begin{eqnarray}
	\sloep=&2\mpla\left( \frac{H'(\phi)}{H(\phi)} \right)^2&=\frac{1}{2\mpla}\frac{\dot\phi^2}{H^2}\\
	\sloet=&2\mpla\frac{H''(\phi)}{H(\phi)}&=-\frac{\ddot\phi}{H\dot\phi}~.
	\label{slowrollparam}
\end{eqnarray}
We have further more added a second order parameter
\begin{equation}
\sloka=2{M_\text{pl}}^2\frac{H'''(\phi)H'(\phi)}{H(\phi)^2}+\sloet^2=\frac{\dddot\phi}{H^2\dot\phi}~.
\label{sloka}
\end{equation}
To leading order they are related to the usual slow-roll parameters by $\ep_H\approx \ep$, $\eta_H\approx \eta-\ep$.


\section{The de Sitter limit}

Since the curvature perturbation is a pure gauge mode in de Sitter space, the action of the comoving curvature perturbation $\zeta$ must be slow-roll suppressed  to all orders in perturbations and vanish in the de Sitter limit. However, this does not explain why the third and fourth order action of $\zeta_n$ are actually suppressed by the same power of the slow-roll parameter. If we want to estimate the slow-roll order of the $n$-point function of $\zeta$, we would overestimate it, if we only assumed that the action of $\zeta$ has to be slow-roll suppressed.

In fact, the slow-roll order of the action is more easily understood in the uniform curvature gauge. In this gauge, it has been shown that the third order action of the inflaton field fluctuations, $S_3(\delta\phi)$, is suppressed by the square root of the slow-roll parameters \cite{Maldacena:2002vr}, while the second, and fourth order actions, $S_2(\delta\phi)$, $S_4(\delta\phi)$, are unsuppressed \cite{Sloth:2006az,Seery:2006vu}. Thus, we may wonder what requires the third order action, $S_3(\delta\phi)$, to be slow-roll suppressed, when there is nothing which forces the second and fourth order actions to be slow-roll suppressed.

However, if there are third order terms of $\delta\phi$ in the action, which are unsuppressed and survive in the pure de Sitter limit, they would indicate an instability of the classical de Sitter vacuum. As an example, we will study a toy unsuppressed third order term.  In the fourth order action there are unsuppressed terms of the type $\dot{\delta\phi}^2\p^{-2}(\p_i\dot{\delta\phi}\p^i\delta\phi)$ etc. Let us assume for a moment that to third order we have similar unsuppressed terms induced by perturbations of the metric. Then the interaction Hamiltonian for the perturbations will take the form
 \beq
 H_I = \int d^3 y a^3\left[\delta\phi(\ddot\phi_c+3H\dot\phi_c+V')+gO(\delta\phi^3)+\dots \right]~,
 \eeq
where the unsuppressed $O(\delta\phi^3)$ toy term could be any operator of the type $\delta\phi^3$, $\dot{\delta\phi}^2\delta\phi$, $\dot{\delta\phi}\p^{-2}(\p_i\dot{\delta\phi}\p^i\delta\phi)$, etc. The dots represents any terms to higher order in slow-roll and perturbation theory, and $g$ is a coupling constant.  
 
The term $O(\delta\phi^3)$ will give a contribution to the tadpole diagram in fig.~(1), which will lead to a one-loop correction to the equation of motion of the classical background field. The tadpole condition yields
 \beq \label{loopeqmot}
 0 =\left<\delta\phi\right> = \ddot\phi_c+3H\dot\phi_c+V' +g\Gamma_t ~,
 \eeq
where $g\Gamma_t$ denotes the amputated tadpole contribution, and gives the one-loop correction to the background equation of motion. In the simplest case of a massless scalar field and with $O(\delta\phi^3)=H\delta\phi^3$, the tadpole contribution would become
 \beq
 g\Gamma_t  = 3gH\left< \delta\phi^2\right> = \frac{3g}{4\pi^2}H^4t~.
 \eeq
In this case, the time-independent de Sitter solution is destabilized by the tadpole. If the toy term $O(\delta\phi^3)$ has a more complicated form involving derivatives, the infrared divergency will be absent\footnote{The actual tadpole was calculated to leading order in slow-roll in ref.~\cite{Seery:2007we}. Note, that to leading order in slow-roll, the IR divergent terms are not present. The IR divergent contributions to the tadpole, which will dominate the tadpole contribution at late times, appears to higher order in slow-roll, and can be found by taking the appropriate infrared limit in the action as in ref.~\cite{Sloth:2006az,Sloth:2006nu}.}, and the tadpole contribution will not grow indefinitely, but rather approach a constant. However, in the case of a massless scaler field in de Sitter, a time-independent tadpole contribution $g\Gamma_t= const.$ will give an effective linear contribution to the potential of the background term, similar to a source term, which will yield the potential of the massless field unbounded from below. In fact, the solution to eq.~(\ref{loopeqmot}), with $V'=0$ and $g\Gamma_t= const.$ at late times is 
 \beq
 \phi_c(t) = -\frac{g\Gamma_t}{3H}t,
 \eeq
which is inconsistent with a time-independent de Sitter solution. Thus, if unsuppressed third order terms were allowed, classical de Sitter space with a massless scalar field would be destabilized.

\begin{figure}[H]
    \begin{center}
        \includegraphics[width=4cm]{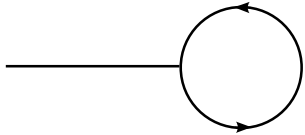}
    \end{center}
    \caption{Tadpole diagram}
    \label{fig:tadpole}
\end{figure}

Since any unsuppressed odd order terms in the action would lead to a non-vanishing tadpole contribution, we conclude that all odd order terms in the action should be slow-roll suppressed, while even order terms are not slow-roll suppressed. We can then extrapolate our results from $n\leq 4$ to any $n$, as shown in table 1. Using the extrapolation in table 1, we can calculate the order of magnitude  of any $n$-point correlation function in single field inflation, as shown in table 2. We can also generalize the nonlinearity parameter $f_{NL}$ to $n$'th order up to a numerical factor of order one
\beq
\left<\zeta^n\right> \approx f_{NL}^{(n)}\mathcal{P}_{\zeta}^{n-1}~,
\eeq 
where $\mathcal{P}_{\zeta}$ is the power-spectrum of comoving curvature perturbations. To second order the generalized nonlinearity parameter coincides with the usual one $f_{NL}^{(3)} = f_{NL}$ for the bi-spectrum, which was initially calculated in ref.~\cite{Acquaviva:2002ud,Maldacena:2002vr}. To third order it coincides with the nonlinearity parameter for the tri-spectrum $f_{NL}^{(4)} = \tau_{NL}$, which was calculated in ref.~\cite{Seery:2006vu}.

\begin{table}[H]
	\centering
	\begin{tabular}{| l ||c|c|c|}
		\hline
		Order&$S(\delta\phi)$&$S(\zeta_n)$&$S(\zeta)$\\
		\hline
		\hline
		2nd&$\mathcal{O}(1)$&$\mathcal{O}(\ep)$&$\mathcal{O}(\ep)$\\
		\hline
                3rd&$\mathcal{O}(\ep^{1/2})$&$\mathcal{O}(\ep^2)$&$\mathcal{O}(\ep)$\\
                \hline
                4rd&$\mathcal{O}(1)$&$\mathcal{O}(\ep^2)$&$\mathcal{O}(\ep)$\\
               \hline
		 \multicolumn{1}{:l:;{4pt/2pt}}{$2n$th}&\multicolumn{1}{c ;{4pt/2pt}}{$\mathcal{O}(1)$}&\multicolumn{1}{c;{4pt/2pt}}{$\mathcal{O}(\ep^n)$}&\multicolumn{1}{c ;{4pt/2pt}}{$\mathcal{O}(\ep)$}\\
		\hdashline[4pt/2pt]
		\multicolumn{1}{:l:;{4pt/2pt}}{$(2n+1)$th}&\multicolumn{1}{:c;{4pt/2pt}}{$\mathcal{O}(\ep^{1/2})$}&\multicolumn{1}{c;{4pt/2pt}}{$\mathcal{O}(\ep^{n+1})$}&\multicolumn{1}{c;{4pt/2pt}}{$\mathcal{O}(\ep)$}\\
		\hdashline[4pt/2pt]
	\end{tabular}
	\caption{Slow-roll order of the action to $n$'th order.}
	\label{tab:system}
\end{table}

\begin{table}[H]
	\centering
	\begin{tabular}{| l || c | c | c |}
		\hline
		$p$& $\left<\delta\phi^p\right>$ & $\left<\zeta^p\right>$& $f_{NL}^{(p)}\approx \left<\zeta^p\right>/\mathcal{P}_{\zeta}^{p-1}$\\
		\hline
		\hline
		2&$\mathcal{O}(H^2)$&$\mathcal{O}(\ep^{-1}H^2)$&$\mathcal{O}(1)$\\
		\hline
                3&$\mathcal{O}(\ep^{1/2}H^4)$&$\mathcal{O}(\ep^{-1}H^4)$&$\mathcal{O}(\ep)$\\
                \hline
                4&$\mathcal{O}(H^6)$&$\mathcal{O}(\ep^{-2}H^6)$&$\mathcal{O}(\ep)$\\
               \hline
		 \multicolumn{1}{:l:;{4pt/2pt}}{$2n$}&\multicolumn{1}{c ;{4pt/2pt}}{$\mathcal{O}(H^{2p-2})$}&\multicolumn{1}{c;{4pt/2pt}}{$\mathcal{O}(\ep^{-p/2}H^{2p-2})$} &\multicolumn{1}{c;{4pt/2pt}}{$\mathcal{O}(\ep^{p/2-1})$}\\
		\hdashline[4pt/2pt]
		\multicolumn{1}{:l:;{4pt/2pt}}{$2n+1$}&\multicolumn{1}{:c;{4pt/2pt}}{$\mathcal{O}(\ep^{1/2}H^{2p-2})$}&\multicolumn{1}{c;{4pt/2pt}}{$\mathcal{O}(\ep^{(1-p)/2}H^{2p-2}))$}&\multicolumn{1}{c;{4pt/2pt}}{$\mathcal{O}(\ep^{(p-1)/2})$}\\
		\hdashline[4pt/2pt]
	\end{tabular}
	\caption{Slow-roll order of the $n$-point functions and generalized nonlinearity parameter.}
	\label{tab:system}
\end{table}

It is easy to verify that the action $S(\delta\phi)$ to any even order, $2n$, in perturbations will be unsuppressed in the slow-roll parameters, as it will contain contributions from $\od{\alpha}{2}^{2n}$, which is unsuppressed in the slow-roll parameters. Similarly, to odd orders $\od{\alpha}{2}$ will always appear in combination with some $\od{\alpha}{n}$ or $\od{\chi}{n}$ to odd order, say $\od{\alpha}{1}$ which is slow-roll suppressed.

As an example we predict that the nonlinearity parameter related to the $5$- and $6$-point function is $f^{(5)}_{NL}=f^{(6)}_{NL}=\ep^2$.

Finally, let us briefly discuss what would be the effect of including gravitational wave modes in the analysis. For gravitational waves the discussion of the de Sitter limit is a little more involved. One can have unsuppressed odd $n$-terms in the action with gravitational waves, $\gamma_{ij}$, of the form $\p_i\delta\phi\p_j\delta\phi \gamma^{ij}$, since they will not contribute to the tadpole of the scalar field fluctuation. However, terms like $\delta\phi \dot\gamma_{ij}\dot\gamma^{ij}$ have to be slow-roll suppressed, because they will contribute to the tadpole with a graviton circulating in the loop. This agrees with the results of ref.~\cite{Maldacena:2002vr}, where the third order action including gravitational waves has been calculated.


\section{Conclusions}

We have calculated the fourth order action of the comoving curvature perturbation, and the third order gauge transformation between the uniform curvature gauge and the comoving gauge. We have shown that the fourth order action of the comoving curvature perturbation is suppressed by one power of the slow-roll parameter, and when terms proportional to the linear perturbation equation is removed by a field redefinition, the relevant action is suppressed by two orders in slow-roll. This is consistent with the previous calculations in the uniform curvature gauge. 

In the uniform curvature gauge, it has been shown that the action of inflaton field perturbations is unsuppressed to fourth order, even though the action is slow-roll suppressed to third order. We argued in the previous section, that to any odd order the action of inflaton field perturbations has to be slow-roll suppressed in order for the classical tree-level de Sitter vacuum to be stable, since otherwise a non-vanishing tadpole contributions would shift the de Sitter solution. Since the even order terms in the action does not contribute to the tadpole, they do not have any destabilizing effect on the background, and are thus allowed to survive the de Sitter limit.

Extrapolating this argument to $n$'th order, we can estimate the slow-roll order of the action to any order, and use it to estimate the $n$-point function of inflaton perturbations to arbitrary order. Although the non-gaussianity from the bi-spectrum is difficult to measure at present, and the higher order sources of non-gaussianity will be even harder to detect, the result gives a useful theoretical insight that can be used for different purposes. If one is interested in possible large non-gaussianities from more exotic models of inflation, it is useful to understand carefully the simplest case of single field slow-roll inflation first. In fact the calculated fourth order action of comoving curvature perturbations can be used to calculate the enhancement of the tri-spectrum due to non-adiabatic effects during inflation \cite{hhjs}.

\overskrift{Acknowledgments}

\noindent M.S.S. would like to thank Paolo Creminelli,  Nemanja Kaloper, Marcello Musso, David Seery and Filippo Vernizzi for useful discussions. In addition we would like to thank Emanuela Dimastrogiovanni for pointing out some important typos in the previous version.

\appendix
\appendix
\section{Auxillary functions}
\label{aux}
\begin{equation}
	\od{\alpha_\phi}{1}=-\frac{\dot\phi_c^2}{2H^2}\zeta_n
	\label{alpha1phi}
\end{equation}
\begin{equation}
	\od{\chi_\phi}{1}=\frac{\dot\phi_c^2}{2H^2}\partial^{-2}\dot\zeta_n
	\label{chi1phi}
\end{equation}
\begin{equation}
	\begin{split}
		\od{\alpha_\phi}{2}=\frac{\dot\phi_c^4}{8H^4}\zeta_n^2-\frac{1}{2H}\partial^{-2}\bigg(&\frac{\dot\phi_c}{H}\partial_i\left[ \frac{\dot\phi_c\dot H}{H^2}\zeta_n-\frac{\ddot\phi_c}{H}\zeta_n-\frac{\dot\phi_c}{H}\dot\zeta_n \right]\partial^i\zeta_n+\frac{\dot\phi_c^4}{4H^4}\dot\zeta_n\partial^2\zeta_n\\
		&+\frac{\dot\phi_c}{H}\left[ \frac{\dot\phi_c\dot H}{H^2}\zeta_n-\frac{\ddot\phi_c}{H}\zeta_n-\frac{\dot\phi_c}{H}\dot\zeta_n \right]\partial^2\zeta_n-\frac{\dot\phi_c^4}{4H^4}\partial_i\partial_j\zeta_n\partial^i\partial^j\partial^{-2}\dot\zeta_n\bigg)
\end{split}
	\label{alpha2phi}
\end{equation}
\begin{equation}
	\begin{split}
		4H\partial^2\od{\chi_\phi}{2}=&-\frac{\dot\phi_c}{H^2}\partial_i\zeta_n\partial^i\zeta_n-V''\frac{\dot\phi_c^2}{H^2}\zeta_n^2+\frac{\dot\phi_c^4}{4H^4}\dot\zeta_n^2-\frac{\dot\phi_c^4}{4H^4}\partial_i\partial_j\partial^{-2}\dot\zeta_n\partial^i\partial^j\partial^{-2}\dot\zeta_n\\
		&-\frac{\dot\phi_c^4}{H^3}\partial_i\partial^{-2}\dot\zeta_n\partial^i\zeta_n-\frac{2\dot\phi_c^3}{H^2}\left( \frac{\dot\phi_c\dot H}{H^2}-\frac{\ddot\phi_c}{H}\right)\zeta_n^2\\
		&-(\frac{3\dot\phi_c^4}{4H^4}\zeta_n^2-2\od{\alpha_\phi}{2})(-6H^2+\dot\phi_c^2)-\left[ \frac{\dot\phi_c\dot H}{H^2}\zeta_n-\frac{\ddot\phi_c}{H}\zeta_n-\frac{\dot\phi_c}{H}\dot\zeta_n \right]^2
	\end{split}
	\label{chi2phi}
\end{equation}
\begin{equation}
	\begin{split}
		\frac{1}{2}\od{ {\beta_\phi}_i}{2}=\partial^{-2}\bigg(&2H\partial_i\od{\alpha_\phi}{2}-\frac{\dot\phi_c^4}{H^3}\zeta_n\partial_i\zeta_n+\frac{\dot\phi_c}{H}\left[ \frac{\dot\phi_c\dot H}{H^2}\zeta_n-\frac{\ddot\phi_c}{H}\zeta_n-\frac{\dot\phi_c}{H}\dot\zeta_n \right]\partial_i\zeta_n\\
		&+\frac{\dot\phi_c^4}{2H^3}\zeta_n\partial_i\zeta_n+\frac{\dot\phi_c^4}{4H^4}\dot\zeta_n\partial_i\zeta_n-\frac{\dot\phi_c^4}{4H^4}\partial_j\zeta_n\partial_i\partial^j\partial^{-2}\zeta_n\bigg)
	\end{split}
	\label{beta2phi}
\end{equation}



\end{document}